# Next Generation Language Resources using Grid

Federico Calzolari[+], Eva Sassolini[°], Manuela Sassi[°], Sebastiana Cucurullo[°], Eugenio Picchi[°], Francesca Bertagna[°], Alessandro Enea[°], Monica Monachini[°], Claudia Soria[°], Nicoletta Calzolari[°]

[+]SNS - Scuola Normale Superiore
Piazza dei Cavalieri 7
56126 PISA - Italy
federico.calzolari@sns.it

[°]CNR - Istituto di Linguistica Computazionale
Area della Ricerca di Pisa, via Moruzzi 1
56124 Pisa, Italy
eva.sassolini@ilc.cnr.it; nella.cucurullo@ilc.cnr.it; picchi@ilc.cnr.it; francesca.bertagna@ilc.cnr.it; enea@ilc.cnr.it;
monica.monachini@ilc.cnr.it; claudia.soria@ilc.cnr.it; glottolo@ilc.cnr.it

**Abstract**
This paper presents a case study concerning the challenges and requirements posed by next generation language resources, realized as an overall model of open, distributed and collaborative language infrastructure. If a sort of "new paradigm" is required, we think that the emerging and still evolving technology connected to Grid computing is a very interesting and suitable one for a concrete realization of this vision. Given the current limitations of Grid computing, it is very important to test the new environment on basic language analysis tools, in order to get the feeling of what are the potentialities and possible limitations connected to its use in NLP. For this reason, we have done some experiments on a module of Linguistic Miner, i.e. the extraction of linguistic patterns from restricted domain corpora.

## 1. Introduction

The growing bulk of textual material available on the Internet is an inexhaustible source of linguistic knowledge. At the same time, the automatic treatment of massive repositories of data requires tools more and more sophisticated and powerful. In the last few years it is perceived as essential to define a general organization and plan for research, development and cooperation in the Language Resources (LR) area, to avoid duplication of efforts, to build on each other results, and to provide for a systematic distribution and sharing of knowledge. The pooling together of these types of requirements made compulsory and almost inevitable the emergence of the idea of the essential infrastructural role that LRs play in HLT.

This requirement of sharing Language Resources and Language Technologies, and of making it possible a real accumulation of knowledge, gives birth, in a natural way, to the idea of exploiting the Grid architecture, i.e. a world-wide distributed network of computers that work in a coordinated way. Grid can be seen as a sort of distributed calculus on geographic scale. By exploiting the computing power of a very high number of processors, this architecture allows not only real time analysis of massive textual corpora but also to distribute several processing requests over several nodes at the same time, thus allowing a parallel search.

The techniques of Grid computing are well-suited for the scientific-experimental field, where the combination of the power of several processors allows obtaining a greater computing power. It remains the question whether Grid can be useful also in other fields, by providing higher flexibility and optimization of the exploitation of systems and resources.

At the moment there are not many groups world-wide that are working towards this direction. Among the few we mention Edinburgh University (Huges, 2001) and Tokyo Institute of Technology (Furui, 2006).

In section 2 we briefly introduce the Grid and the Italian Grid in particular, to which our institute (ILC) belongs; in section 3 we describe the Linguistic Miner system, an existing software with components that can run over the Grid (at the moment for experimental purposes); section 4 is the core of the paper describing our first experiments in using the Linguistic Miner on the Grid and providing first results; we then give our perspectives for future work in section 5.

## 2. What is the Grid

Grid computing is a form of distributed computing that involves coordinating and sharing computers, applications, data, storage, and network resources across dynamic and geographically dispersed organizations. The Grid aims ultimately to turn the global network of computers into one vast computational resource. Grid is in fact a sort of low-cost world-wide shared super-cluster, designed to deal with the increasing computational and storage load, far beyond a single institute or group computing capability.

The system is based on multiple Virtual Organizations (VO), a set of applications, users and rules governing resources sharing. Grid allows to analyze in a few minutes a data volume of several TeraBytes (up to PetaBytes), distributing the computational load between multiple CPUs in different locations. Job submission is managed by the Grid user, whereas failure is handled by the Grid middleware. The best solution in terms of proximity, power and storage required, is provided by Grid infrastructure, using the Resource Broker information.

### 2.1. The Italian Grid

The Italian Grid is part of a European Grid that is being built under the EGEE (Enabling Grid for e-sciencE

in Europe) project, formally LCG (LHC - Large Hadron Collider - Computing Grid) project that aims to establish a Grid infrastructure for European science (see www.eu-egee.org for more details). Nine federations (called ROCs) participate into the European grid. Each federation is composed by one or more European states plus the Russia.

The INFNGrid project started in 1999, developing the first Italian Grid, based on GARR, the Italian research network. The INFNGrid counts more than 20 sites among the most important Italian universities and, although primarily focused on the development of computing infrastructures for physics, it has been, since the beginning, open to other fields of research (bio-medicine, earth observation, etc.) and to industry. It is now a successful example of collaboration between physicists, software engineers, computer professionals, computer scientists.

ILC is part of LCG-INFNGrid production network since 2005 as one of the first not INFN grid nodes in Italy, where it represents the first computational linguistics institute in Europe.

## 3. The Linguistic Miner

The Linguistic Miner (Picchi *et al.*, 2004) is a project started in 2003 with the aim of developing a framework for the automatic extraction of linguistic knowledge from very large amounts of texts in Italian language (from different sources and in different formats) to be exploited in many ways and also with didactic, editorial and cultural aims. The basic idea of the Linguistic Miner is that big and representative textual material in different language fields (wrt genres, contexts of usage, etc.) are close to the linguistic reality of a language.

The last years, there is an increase of the techniques of knowledge extraction (*data mining*) exploiting implicit and explicit relations among textual data as opposed to simple retrieval operations (*data retrieval*). The initial phase of the project was dedicated to the gathering of corpora of Italian texts and it produced a repository (a "mine") of around 200 millions words together with a systematic (automatic) topic classification of texts.

This was achieved by exploiting procedures for the upgrade and augmentation of textual data in the "mine" and for the automatic acquisition from the Web (*spiders*), both with periodic updating and by means of user-defined paths. In the case, for instance, of periodic navigation of a web site, a user can decide which links to visit and insert them into a script that will guide the spidering program during subsequent access.

The second phase of the project consisted in the automatic encoding of the textual material, by using modules of the PiSystem (Picchi, 1994), an integrated framework for the treatment of textual and lexical material, where the most important module is the DBT (Data Base Testuale, *Textual Data Base*), the module for the linguistic analysis. The most effective procedures for further analysis of texts are POS tagging and lemmatization, which have been performed over 90% of the whole repository.

The focal point of the entire project is the phase of exploitation of the textual data in the "mine", which provides results such as extraction of concordances of word occurrences, single lemmas, specific multiwords, individuation of structured left and right-neighbour co-occurrences, pattern matching procedures.

### 3.1. Specific applications

The extraction of linguistic patterns from the reference corpus was the component chosen as a test application in the Grid framework. It represents an important instrument for the analysis of language, allowing not only the search of information and the verification of linguistic hypotheses, but also the construction of a data bank of what was searched, analysed, and extracted: for example, the results of the extraction of domain terms from a given corpus may be used to create a terminological dictionary; syntactic patterns can be stored and subsequently analyzed to form verb classes, etc.

The basic idea is the incremental building of a set of repositories of integrated linguistic data containing the memory of all the info and linguistic notions previously gathered. The system can be seen as a set of integrated modules cooperating to acquire, code and analyze the results. In this way we obtain a huge data bank of texts marked by different users for specific purposes, available for further, more sophisticated processing steps. This property makes the Linguistic Miner a very interesting tool for our experiment of using Grid Technology in the field of Human Language Technology.

### 3.2. LM on Grid

Due to the high self-incrementing power of LM system, very soon we had to deal with huge quantities of data to process. We borrowed from Grid computing approach the idea of (i) exploiting the unprecedented power represented by the number of computers distributed over the planet and the existence of networks to interconnect them, in order to carry out time-consuming task and (ii) allocating pre-annotated and classified language resources over different nodes available on the grid. In the grid approach, it does not matter where specific routines are executed, rather how synchronous they are.

One of the principles underlying Grid computing is that it is not important what processor(s) will execute the specific sequence of instructions because the overall Grid architecture will individuate the "most suited" processor among those available and it will execute the code.

An easy, transparent, coordinated and secure access to resources is enabled by procedures and technologies available in the VO a user belong to.

The Italian component INFNGrid (see section 2) allowed us to select *n* nodes candidate on the grid where to allocate our resources. The potentialities offered by such infrastructure will dramatically alter the face and the nature of the model underlying LM with its applications. In this context, the system can support, e.g., the extraction of linguistic patterns in the form of services in a distributed and controlled environment. From another perspective, LM implementation on a Grid platform forces technologies available in LM to be made interoperable with grid technologies. The development environment used for LM is radically different from that adopted for the development of applications running under Grid, i.e. S.O. Unix.

Within LM, those technologies are individuated which are (i) not heavily dependent software libraries available only on S.O. Windows, (ii) easily portable in the Unix environment, and (iii) particularly fit for being partitioned and executed over multiple systems. In any case, the task "LM-to-Grid", i.e. porting to Grid environment any of the applications available in LM, always implies to write suitable conversion routine(s) in order to ensure compliance to the grid "regulations", i.e.:

a) Once a resource is released by a process, it is no longer possible to access it for subsequent processing phases. It is not allowed to access many times the same resource data as normally happens with LM.
b) Once the elaboration is launched, this will result in an output file. It is not possible to produce intermediate files on which to perform e.g. sorting operations as currently happens with LM.

At present, modularity of programs has been a winner strategy in LM, enabling the combination of different modules for different computations. In Grid environments those modules have been re-written and deployed from scratch in order to obtain the expected results.

Given those premises, in order for grid computing to become a reality it is crucial the role the software tools play. Algorithmic efficiency and performance is to be evaluated no more in terms of execution speed but as concerns the capability of dealing with the specific requirements. In other words, the availability of a huge power of parallel computation, can direct the researcher, who has to develop a new application, to opt for intrinsically parallel algorithms, since the beginning. The task of building a grid-enabled version of our LM and porting already existing software to the new Grid paradigm is greatly harder, as in the case at hand.

## 4. LM applications in the Grid environment

The assumption driving us towards the Grid is that the data collected in the Linguistic Miner framework represent a stable set that contains all the necessary basic information (lemmatized texts with functional codes) and is subject to a continuous growth. These data are an inexhaustible source of research but, at the same time, they may become difficult to be locally handled and processed.

One of the applications available in LM is represented by pattern-matching rules for searching information in free text[1]. We decided to experiment a Grid-enabled version of this "linguistic pattern" extraction module. As already explained in section 3.2, this implementation implies not only developing conversion procedures of the module to comply with the requirements of the Grid environment, but also adopting a radically different approach in managing and accessing data. The present experiment of LM-to-Grid, to be considered just as a case study preparatory of future work, consists of two distinct aspects:

a) The creation of an *ad-hoc* module for the extraction of noun-adjective and noun-noun sequences that exploits rules based on already tested linguistic patterns;
b) The distribution of the extraction process and of four lemmatized corpus pertaining to the domain of medicine, agriculture, history and natural environment, on many Grid LCG nodes, according to foreseen security procedures (data redundancy in case of downtime of one of the nodes).

The processing output consists in text files composed by ordered lists of lemmas together with the associated adjective (or noun).

### 4.1. Analysis of results

From a first analysis of results it is clear how the management of this considerable quantity of data represents a problem: without a proper strategy it would be easy to waste its value and peculiarities. In a relatively little domain such as the agriculture domain the noun-adjective lists are huge and have to be filtered.

We decided to impose a minimal frequency limit for the noun to be retrieved. In the specific case, a lower threshold was established, that is bound to be incremented as long as the data volume increases. The algorithm has been studied to retrieve also those nouns that, even if with frequency lower than the established threshold, are followed only by a particular adjective. Some of these pairs can be interpreted as multiword expressions and also extracted as hyponyms of the head of the phrase to be used for a further development towards the construction of thesauri and ontologies. Our experiment goes in the direction of enabling us to use the Web as a big linguistic corpus and paves the way towards the establishment of an observatory of the dynamic evolution of the language.

By exploiting the Grid, we extracted terminological sets relative to some important domains, like the "protection of the environment" and medical ones, etc. We obtained the expected results, as we can observe by looking at the examples reported in Figure 1, where the analysis of the newspaper articles makes emerge the terminology used in the medical field. What is quite interesting, in our perspective, is that the Grid can be seen as a mean to achieve the analysis of data that are dynamically and constantly acquired and to monitoring the diachronic evolution of the language. From Fig. 1, we can see that one of the most frequent complex terms is represented by the noun-adjective pair *mucca pazza* ("mad cow"), that was, in a given period, one of the most discussed topics.

Although the data used for this experiment amount to a few hundreds of compressed MB (corresponding to about 40 million words), the distribution of the job load over the computational grid allows to reduce the processing time by n times, where n is the number of datasets to be analyzed, or, in other words, the sub-corpus (already analyzed and tagged by LM) to which the extraction job is applied. If we hypothesise huge quantities of data, such as a high number of corpora, and/or a high computational load required for the linguistic analysis, the Grid environment undoubtedly offers an advantage in terms of required time and costs.

---

[1] This application was e.g. exploited in order to extract lists of bigrams from 48,000 sentences of the Italian Cassazione Court over the period 1990-2001 to be used as a base for the classification of their content (Sassi and Cinini, 2004).

| 04839 | [MALATTIA] | ACUTO 23, ALLERGICO 26, ARTEROSCLEROTICO 82, CARDIACO 118, **CARDIOVASCOLARE 778**, CAUSATO 28, CELIACO 15, COLLEGATO 12, CONCLAMATO 18, CONGENITO 27, CONTAGIOSO 17, **CORONARICO 306**, CRONICO 170, CURABILE 15, CUTANEO 15, **DEGENERATIVO 145**, DIVERSO 13, DOVUTO 17, ENDEMICO 11, ENDOCRINO 23, EREDITARIO 132, ESANTEMATICO 143, **GENETICO 360**, GRAVE 87, INCURABILE 20, INFETTIVO 613, ............ ............ |
|---|---|---|
| 01670 | [MUCCA] | MALATO 24, **PAZZO 1593**, ... |

**Fig. 1 – list of the most frequent nouns**

## 5. Future Perspectives

The Grid environment has produced the desired results:

i) reduction of the processing time proportionally to the number of the nodes on which the calculus is distributed;
ii) huge storage capacity;
iii) data redundancy;

These results are obtained without any additional cost for the final user.

The Grid environment is a good candidate to make available both the linguistic resources and the tools for their processing to a credited, specialized yet broader public. This hypothesis would imply greater attention to data protection and accessibility to resources; however, this is already made feasible by existing tools such as accounting methods and protected data access only by users owning a certificate issued by their VO. The results obtained encourage future developments; in the particular case at hand, a further step is represented by the standardization of linguistic patterns and by the possibility for the user to exploit scripts to customize the query.

From a more strategical point of view, we noticed that a new paradigm of R&D in LRs and LT is emerging, pushing towards the creation of open and distributed linguistic infrastructures for LRs and LT, based on sharing LRs and tools. It is urgent to create a framework – both technological and organizational – that enables controlled and effective cooperation of many groups on common tasks, adopting the paradigm of accumulation of knowledge so successful in more mature disciplines, such as biology and physics. This implies the ability to build on each other achievements, to merge results, and to have them accessible to various systems and applications. This is the only way to make a clear leap forward. This means emphasizing interoperability among various initiatives related to LRs, LT and knowledge bases. To mention just one example, more and more initiatives are arising aimed at achieving international consensus on annotation guidelines: to be able to merge diverse linguistic annotation efforts, produce a set of coherent, integrated, comprehensive linguistic annotations to be readily disseminated throughout the community. Our claim is that this can be facilitated by application of Grid technology to tackle the problems of processing extremely large quantities of "facts and their relations", of development of unprecedented large-scale annotated LRs, and of their dynamic linking across many different sources. A difficulty and a challenge will be how to coordinate different information sources.

Moreover, we end by noting how in the domain of Italian broadcasting the citation of publications is a common strategy to propose statistical data relative to frequency of word use in order to define or exemplify the behaviour of institutions or public people. This remark is to show how linguistic statistics is already commonplace and can be used and thus required to offer a matter of reflection to a wider public than scholars of a restricted discipline.